\documentclass[prd,twocolumn,nofootinbib]{revtex4}
\usepackage{dblfloatfix}
\usepackage{dcolumn}
\usepackage{graphicx}
\usepackage{epsfig} 
\usepackage{amsmath}
\usepackage{amsfonts}
\usepackage{amssymb}
\usepackage{color}
\usepackage{float}
\usepackage[caption = false]{subfig}

\begin{document}

\title{Consistency of the growth rate in different environments with
  the 6dF Galaxy Survey: \\ measurement of the void-galaxy \&
  galaxy-galaxy correlation functions}

\author{I. Achitouv$^{\dagger\dotplus}$\footnote{E-mail:iachitouv@swin.edu.au}}
\author{C. Blake$^{\dagger\dotplus}$}
\author{P. Carter$^\star$}
\author{J. Koda$^\ddagger$}
\author{F. Beutler$^\star$}

\address{$^{\dagger}$Centre for Astrophysics \& Supercomputing,
 Swinburne University of Technology, P.O. Box 218, Hawthorn, VIC
  3122, Australia \\
   $^{\dotplus}$ARC Centre of Excellence for All-sky
 Astrophysics (CAASTRO) \\
 $^\star$  Institute of Cosmology \& Gravitation, Dennis Sciama Building, University of Portsmouth, Portsmouth, PO1 3FX, UK \\ 
 $^\ddagger$  Department of Mathematics and Physics, Universita Roma Tre, Via della Vasca Navale 84, Rome 00146, Italy }

\begin{abstract}
We present a new test of gravitational physics by comparing the growth
rate of cosmic structure measured around voids with that measured
around galaxies in the same large-scale structure dataset, the
low-redshift 6-degree Field Galaxy Survey.  By fitting a
Redshift Space Distortion model to the 2D
galaxy-galaxy and void-galaxy correlation functions, we recover growth
rate values $f \sigma_8 = 0.42\pm 0.06$ and $0.39 \pm 0.11$,
respectively.  The environmental-dependence of cosmological statistics
can potentially discriminate between modified-gravity scenarios which
modulate the growth rate as a function of scale or environment and
test the underlying assumptions of homogeneity and isotropy.
\end{abstract}

\maketitle

\section{Introduction}

Galaxy peculiar velocities are a powerful probe of gravitational
physics.  They are sourced by virialized motion within halos and the
overall bulk-flow motions due to gravitational interactions, leading
to the mass assembly of halos.  Although direct measurement of galaxy
peculiar velocities is challenging, their correlated effect is
imprinted in the clustering of matter through Redshift Space
Distortion (RSD), allowing us to determine the linear growth rate of
structure.  This quantity describes the growth of matter perturbations
through cosmic evolution, containing critical information on cosmic
expansion and gravitational physics.

\medskip

For standard General Relativity (GR), in homogeneous and isotropic
cosmologies, the growth rate in linear perturbation theory does not
depend on the comoving spatial scale \cite{Peebles} and can be
approximated by $f \sim \Omega_m(z)^\gamma$ where $\Omega_m$ is the
matter density parameter at redshift $z$, and $\gamma$ is a constant.
For a $\Lambda$CDM Universe $\gamma \sim 0.55$, independently of scale
and environment.  This would not be the case for different
cosmological scenarios.  For instance, inhomogeneous models of dark
energy can lead to patches of clustered dark energy
(e.g. \cite{Vernizzi2015}, \cite{DuttaMaor2007}) which will have
different expansion histories, or certain models of modified gravity
such as $f(R)$ \cite{Hu_Sawicki_2007} rely on the Chameleon effect
\citep{Khoury_Weltman_2004} that suppresses the gravitational force in
under-dense environments.  These theories would naturally lead to an
environmentally-dependent growth rate and possibly a breakdown of the
cosmological isotropy of our universe. As pointed out in
\cite{Caietal2016}, the scale on which the environment is defined is
important.  For very large under-dense regions, the effective
cosmological parameters are expected to be different to the
global-averaged parameters, but the quantification of this critical
scale can also serve as an interesting test for departures from
Einstein gravity.

\medskip

A simple test of this physics is to compare the growth rate around
cosmic voids to that inferred from galaxy clustering. In fact,
non-linear dynamics are expected to be reduced in cosmic voids
compared to galaxy clustering in overdense regions \cite{Hamaus2016}.
Hence cosmic voids can potentially provide powerful tests of
cosmology, for instance using the integrated Sachs-Wolfe effect
\cite{SachsWolfe1967} (e.g. \cite{Granett2008}), the Alcock-Paczynski
test \cite{Alcock-Paczynski1979} (e.g. \cite{LavauxWandelt2012}) or
void abundance and density profile
(e.g. \cite{ANP,IA_voidfinder,ABPW,Clampittetal2013,Zivicketal2015}).

\medskip

In this work we test the consistency of the growth rate with
environment using RSD measurements around voids and galaxies in the
\textit{6-degree Field Galaxy Survey} (6dFGS)
\cite{Jones2004,Jones2006}, a low-redshift large-scale structure
dataset.  There are several advantages to performing these tests near
$z=0$.  First, cosmic expansion is dominated by dark energy, hence a
measurement of the growth rate around cosmic voids is a particularly
interesting test of dark energy clustering.  Second, the impact of the
Alcock-Paczynski effect at $z=0$ is minimal, such that our
measurements have little sensitivity to the assumed cosmology.  Third,
low-redshift surveys such as the 6dFGS have a much higher galaxy
number density than high-redshift surveys, enabling a
higher-resolution measurement of the density field.  This is
particularly important for identifying voids in an unbiased fashion.
Finally, the 6dFGS also contains a set of direct galaxy peculiar
velocity measurements derived using fundamental-plane distances
\cite{Springob2014}.  Although we don't use these measurements in the
present work, they offer interesting opportunities for future
investigation.

\medskip

The measurement of the growth rate using RSD in galaxy clustering has
been previously investigated for many datasets including the 6dFGS
\cite{Beutler6dF}, the 2dF Galaxy Redshift Survey (2dFGRS)
\citep{Peacock2001,Hawkins2003, Cole2005}, the Sloan Digital Sky
Survey (SDSS) \citep{Tegmark2006}, the WiggleZ Dark Energy Survey
\cite{Blake2011}, the Baryon Oscillation Spectroscopic Survey (BOSS)
\citep{Reid2012,Reid2014,Beutler2014} and the VIMOS Public
Extragalactic Redshift Survey (VIPERS) \cite{delaTorre2013}.  These
measurements have shown a general consistency with the $\Lambda$CDM
cosmological model, up to a $2.5\%$ precision, albeit in some cases
showing tension with the predictions of the latest Cosmic Microwave
Background measurements \cite{Planck_cosmo}.  However, the measurement
of the growth using RSD in void-galaxy clustering has not been widely
investigated, although \cite{Hamaus2016} and \cite{Adam_viper}
recently reported measurements using the BOSS-CMASS sample and VIPERS,
respectively.  However, none of these studies has explored the
consistency of the growth rate in different environments using the
same galaxy survey.

\medskip

This paper is organized as follows: in section \ref{sec2} we describe
the model we use to fit the measurement of the galaxy-galaxy and
void-galaxy correlation functions.  In section \ref{sec3} we test
these models using mock catalogues.  In section \ref{sec4} we apply
our framework to the 6dFGS data and deduce constraints on the growth
rate in different environments, and we conclude in section \ref{sec5}.

\section{Models for the 2D correlation functions}
\label{sec2}

The peculiar velocities of galaxies, $\mathbf{v}$, due to the local
gravitational potential, result on small scales in random motions of
galaxies within a group.  By measuring galaxy positions in redshift
space, we can observe the well-known `Finger-of-God' (FoG) effect.  On
large scales, the bulk flow (coherent infall/outflow in
overdense/underdense regions) is responsible for an overall coherent
distortion known as the `Kaiser effect' \cite{Kaiser1987}.

\medskip

The mapping of the position of a galaxy from real space
$\mathbf{r}=(x,y,z)$ to its position in redshift space $\mathbf{s}$ is
given by:
\begin{equation}
\mathbf{s} = \mathbf{r} + \frac{(1+z) v_p(\mathbf{r})}{H(z)}
\mathbf{u_r} ,
\label{RSDeq}
\end{equation}
where $\mathbf{u_r}$ is the unitary vector along the line of sight,
$v_p \equiv \mathbf{v}.\mathbf{u_r}$ and $H(z)$ is the Hubble
parameter at redshift $z$.  On large scales, where the matter
overdensity grows coherently \cite{Kaiser1987,Hamilton1992}, linear
perturbation theory implies that $\triangledown . \mathbf{v} \propto
-f \delta_m$ where $\delta_m$ is the matter density contrast and the
linear growth rate of perturbations $f$ is defined as:
\begin{equation}
f \equiv \frac{d\ln \delta_m(a)}{d\ln a} .
\end{equation}
We need to relate the observed galaxy overdensity, $\delta_g$, to the
matter density contrast, which we accomplish using a linear bias $b
\equiv \delta_g/\delta_m$, which is independent of scale in the linear
regime.

\medskip

In what follows we use the notation $\sigma$ for the component of
galaxy-galaxy or void-galaxy separation perpendicular to the line of
sight, and $\pi$ for the component parallel to the line of sight.  For
both the galaxy-galaxy and the void-galaxy correlation functions, the
random small-scale component of the peculiar velocity can be described
by convolving the correlation function with a pairwise velocity
distribution \cite{Peebles}.  The latter is often modelled as a
Gaussian or Lorentzian distribution; we consider both choices in our
analysis.

\subsection*{The galaxy-galaxy correlation function}

The redshift-space 2D correlation function due to the coherent bulk
flow of peculiar velocity can be described by
\cite{Kaiser1987,Hamilton1992}:
\begin{equation}
\xi^l(\sigma,\pi) = \xi_0(s)P_0(\mu) + \xi_2(s)P_2(\mu) +
\xi_4(s)P_4(\mu) ,
\label{poly}
\end{equation} 
where $P_l(\mu)$ are Legendre polynomials and $\mu \equiv
\cos(\theta)$ is the angle between the separation vector and line of
sight.  In the linear regime \cite{Kaiser1987},
\begin{align*}
\xi_0(s) & = (1 + \frac{2}{3} \beta + \frac{1}{5} \beta^2) \times b^2 \xi(r) \\
\xi_2(s) & = (\frac{4}{3} \beta + \frac{4}{7} \beta^2) \times b^2 \left( \xi(r) - \bar{\xi}(r) \right) \\
\xi_4(s) & = \frac{8}{35} \beta^2 \times b^2 \left( \xi(r) + \frac{5}{2} \bar{\xi}(r) - \frac{7}{2} \bar{\bar\xi}(r) \right) ,
\end{align*}
where $\beta = f/b$, the real-space matter correlation function is
$\xi(r)$, and
\begin{equation}
\bar{\xi}(r) = (3/r^3) \int_{0}^{r} \xi(y) y^2 dy \nonumber
\end{equation}
\begin{equation}
\bar{\bar \xi}(r) = (5/r^5) \int_{0}^{r} \xi(y) y^4 dy \nonumber
\end{equation}
Including our model for small-scale random motions, the total 2D
correlation function in redshift space is given by \cite{Peebles}
\begin{equation}
\xi_{gg}(\sigma,\pi) = \int \xi^l(\sigma,\pi-\frac{v}{H_0}) P(v) dv ,
\label{theogg}
\end{equation}
where $P(v)$ is the probability distribution of the random pairwise
motions.  In what follows we model the matter clustering using the
non-linear power spectrum from \textit{CAMB} (halofit)
\cite{Lewis_camb} and Fourier transform it to obtain the non-linear
matter correlation function $\xi(r)$ in Eq.\ref{theogg}.  We adopt a
fiducial cosmology matching that of Mocks A described below: a flat
WMAP 5-year cosmology \cite{wmap} ($\Omega_m=0.26$, $h=0.72$,
$\sigma_8=0.79$, $n_s=0.963$, $\Omega_b=0.044$).

\subsection*{The void-galaxy correlation function}

The previous effects of the peculiar velocity also apply to the
void-galaxy correlation function and we have \cite{Peebles}:
\begin{equation}
\begin{split}
\xi_{vg}(\sigma,\pi) &= \int (1+\xi_{vg}^{1D}(y)) \times \\ &
P\left(v-v_p(y)\left[\left(\pi-\frac{v}{H_0}\right)/y\right]\right)
dv - 1 ,
\label{theovg}
\end{split}
\end{equation}
where $\xi_{vg}^{1D}$ is the angle-averaged void-galaxy correlation
function in real space and $y = \sqrt{\sigma^2+(\pi-v/H_0)^2}$.

\medskip

We calibrate the model using the real-space void-matter
cross-correlation $\xi_{v-\rm DM}(r)$ measured from N-body simulations
(see section \ref{sec3}) as our $\Lambda$CDM template, such that
including the linear bias factor
\begin{equation}
\xi_{vg}^{1D}(r) = b \; \xi_{v-\rm DM}(r) .
\label{xibias}
\end{equation}

\medskip

For coherent outflow motion, at linear order, the peculiar velocity can
be expressed as \cite{Peebles}:
\begin{equation}
v_p(r) = -\frac{1}{3} H_0 r \Delta(r) f ,
\label{vp_r}
\end{equation}
where $\Delta(r)$ is the average integrated density contrast around
voids.  For spherical voids we have
\begin{equation}
\Delta(r) = \frac{3}{r^3} \int_{0}^{r} \xi_{v-\rm{DM}}(y) y^2 dy .
\end{equation}

\subsection*{The pairwise velocity distribution}

In this work we will consider two models $G$ and $L$ to describe the
pairwise velocity distribution $P(v)$ in Eq.\ref{theogg},\ref{theovg}:
\textit{model G} will use a Gaussian distribution given by
\begin{equation}
P(v) = \frac{1}{\sqrt{2\pi\sigma_v^2}}\exp{\left[ -\frac{v^2}{2\sigma_{v}^{2}}\right] } ,
\label{Geq}
\end{equation}
while \textit{model L} will use a Lorentzian distribution (in Fourier
space) which corresponds to convolution by an exponential distribution
in configuration space:
\begin{equation}
P(v) = \frac{1}{\sqrt{2\sigma_{v}^2}}\exp{\left[ -\frac{\sqrt{2}\vert
      v\vert} {\sigma_v}\right] } ,
\label{Leq}
\end{equation}
where $\sigma_v$ is the standard deviation of the peculiar velocity.
Our model hence neglects the scale dependence of $\sigma_v$
\citep{Hamaus_simuanalysis, Koppetal2016, IA_voidfinder}.

\medskip

A Gaussian distribution of peculiar velocities is often assumed for
the random motions which result from halo relaxation.  However,
numerical studies (e.g. \cite{Takashi}) have shown that a Lorentzian
distribution can provide a better empirical description of the
distribution of peculiar velocities which might result from a
superposition of different-mass haloes.

\subsection*{Summary of the variables}

Our model hence consists of 3 parameters for both the galaxy-galaxy
and void-galaxy correlation functions: the linear bias $b$ which
enters into Eq.\ref{poly},\ref{xibias}, the standard deviation of the
peculiar velocity $\sigma_v$ that enters into Eq.\ref{Geq},\ref{Leq},
and the linear growth rate $f$ that is part of
Eq.\ref{poly},\ref{vp_r}.  We note that, in the linear-theory
approximation, the fitted values of $f$ and $b$ are degenerate with
the assumed normalization of the matter power spectrum, $\sigma_8$.
We reflect this degeneracy by presenting our results in terms of the
normalized variables $f \sigma_8$ and $b \sigma_8$.

\subsection*{{Remarks on the models}}

The RSD models we use in our study, whilst commonly-adopted in the
literature, greatly simplify the non-linear physics which will be
present on these scales.  For example, galaxy bias generally exhibits
non-linear, non-local, scale-dependent and stochastic properties
\cite{Desjacques2016} and the galaxy pairwise velocity dispersion may
be scale-dependent or non-Gaussian \cite{Hamaus_simuanalysis,
  Koppetal2016, IA_voidfinder}.  However, in the following section we will use mock
catalogues to demonstrate that, at the level of statistical precision
of the 6dFGS dataset, these simple models are sufficient to extract
unbiased estimates of the growth rate from both the galaxy-galaxy and
void-galaxy correlations.  Many studies have confirmed this conclusion
through comparison with more sophisticated models
(e.g. \cite{Beutler6dF}, in the context of 6dFGS).  More accurate
modelling of RSD is a significant challenge for upcoming galaxy
surveys with much greater statistical precision such as \textit{Euclid}
\cite{EUCLIDsurvey}.

\section{Tests on mocks}
\label{sec3}

In order to test our analysis pipeline and the limitations of our
models, we measured the growth rate in two sets of mock catalogues.
\textit{Mocks A} are flat-sky mocks with no survey selection function
applied, for which we possess the full set of dark matter and halo
information.  We used these mocks to model the extraction of galaxy
voids from a volume-limited observational sample and the fitting of
the void-galaxy correlation function.  \textit{Mocks B} are
curved-sky mocks which incorporate the full 6dFGS selection function
via detailed halo-occupation modelling.  Although we do not have the
dark matter information to allow tests of the void sample, we used
these mocks to test the fitting of the galaxy-galaxy correlation
function to the flux-limited observational sample. We summarize the
creation of these two sets of mocks below.

\subsection*{Mocks A: volume-limited samples}

To generate Mocks A, we used a sample of dark matter particles and
halos from the DEUSS simulations \cite{DEUSS}.  These simulations were
run for several scientific purposes, as described in
\cite{Alimi,Rasera,Achitouv} and are freely available.  The
simulations were carried out using the RAMSES code \cite{RAMSES} for a
$\Lambda$CDM model calibrated to the WMAP 5-year cosmological
parameters \cite{wmap}.  We used the $z=0$ output of a simulation
generated in a $648^3 \, h^{-3}$ Mpc$^3$ box using $2048^3$ particles.

\medskip

In section \ref{sec4} we will extract galaxy voids from a
volume-limited sample of 6dFGS galaxies.  We built a series of 20 dark
matter ($b=1$) catalogues approximately matching the number density
and volume of this sub-sample, by randomly-selecting $N_p = 15000$ DM
particles a box of side-length $140 \, h^{-1}$ Mpc.  We also built 20
biased galaxy mocks by sub-sampling $N_h = 15000$ halos identified
with the Friend-of-Friends (FoF) algorithm with linking length $0.2$,
selecting the most massive haloes in order to approximately mimic the
6dFGS selection.  Finally, in order to simulate the RSD we used the
flat-sky approximation and shifted the positions of the DM particles
and halos according to Eq.\ref{RSDeq}, using their peculiar
velocities.

\medskip

We note that, when generating these mocks, it is important to match
the DM and halo number density to the galaxy dataset in order to avoid
introducing a bias in the identification of voids between the mock and
the real dataset.  For instance, in \cite{Caietalbias}, the authors
show that the density profile of voids is sensitive to the resolution
of the simulation.

\subsection*{Mocks B: selection-function samples}

In section \ref{sec4} we will use the magnitude-limited 6dFGS sample
to measure RSD from the galaxy-galaxy correlations.  We therefore
supplemented Mocks A with a second simulation set, Mocks B, which
provided a more accurate curved-sky modelling of the survey selection
function and redshift-dependence of the galaxy bias.

\medskip

We built Mocks B from the COLA N-body simulations introduced by
\cite{Koda2016}, using a modified version of the pipeline created by
\cite{Beutler2016} to construct BOSS and WiggleZ mocks.  In brief, we
first fit the central and satellite galaxy halo occupation
distribution of the 6dFGS galaxy sample as a function of luminosity
\cite{Beutler2013}.  By calibrating the luminosity-redshift relation,
we defined the redshift-evolution of the HOD.  Through careful
comparison of the projected and 3D clustering of the mock and data
sample, we iterated the HOD parameters to produce the closest possible
match.  We then applied peculiar velocities along the line-of-sight,
and sub-sampled the resulting distribution with the 6dFGS angular
selection function \cite{BeutlerBAO}.  These mocks will be presented
in more detail by \cite{Carter2017}.

\subsection*{Void-finding in Mocks A}

In our analysis we identified voids with radius $R_v = 20 \, h^{-1}$
Mpc using the void finder developed by \cite{IA_voidfinder}.  This
radius is chosen as a compromise between being small enough to obtain
sufficient voids for an accurate measurement of the void-galaxy
correlation function, but being large enough to select genuinely
underdense patches of matter \cite{Sutter_smallvoids}.

\medskip

\begin{figure}[ht]
\centering
\includegraphics[scale=0.35]{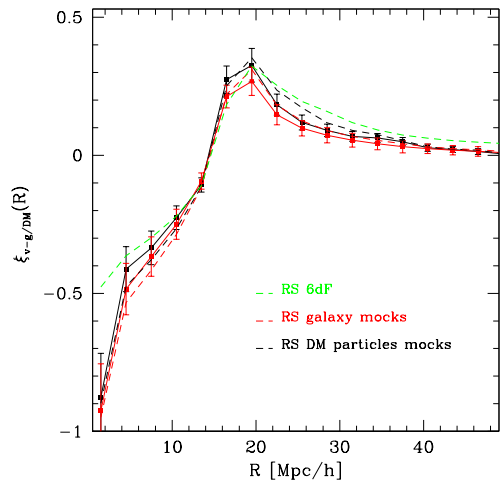}
\caption{Measurement of the 1D void-DM and void-galaxy correlation
  functions.  The error bars show the 1-$\sigma$ standard deviation
  computed using the mock catalogues.  The solid and dashed lines show
  measurements in real-space and redshift-space, respectively, for the
  DM particles (black lines), the haloes (red lines) and the 6dFGS
  galaxies (green line, only available in redshift-space).}
\label{Fig1}
\end{figure}

This void finder uses density criteria to identify voids with the
characteristic profile illustrated by Fig.\ref{Fig1}.  For each of the
candidate void positions, which are picked at random, the algorithm
first requires that the overdensity $\delta$ is below a threshold in
two central bins, $\delta(R_0) < \delta_1 = -0.9$ and
$\delta(R_0+\Delta R) < \delta_2 = -0.8$, where $R_0 = 0.5 \, h^{-1}$
Mpc and $\Delta R = 1 \, h^{-1}$ Mpc.  The third condition ensures a
ridge of the void profile by requiring that $\delta(R_v-\Delta R) <
\delta(R_v)$ and the fourth condition controls the amplitude of the
ridge by requiring that $\delta(R_v) > \delta_3 = 0$.

\medskip

We used 10 times the number of candidate positions as tracers,
producing a sample of $\sim 300$ voids for each Mock A, which is
similar to the number density of voids we find by applying the same
algorithm to the volume-limited 6dFGS sub-sample.  We note that about
half these voids have some portion of overlap; this does not affect
our analysis because overlap does not change the radial density
profile \cite{IA_voidfinder}, and any covariance between overlapping
voids is already encoded in the measurement scatter between mocks.
\subsection*{1D matter-void cross-correlation function}

We measured the void-tracer cross-correlation functions using the
Landy-Szalay estimator:
\begin{equation}
\xi_{vg}(R) = \frac{N_{rg}N_{rv}}{R_vR_g} \left( \frac{D_vD_g}{N_gN_v}
- \frac{D_gR_v}{N_gN_{rv}} - \frac{D_vR_g}{N_vN_{rg}} \right) + 1 ,
\label{LSeq}
\end{equation}
where $D_vD_g$ is the number of data void-galaxy pairs, $R_v R_g$ the
random void-galaxy pairs and $D_{g/v}R_{g/v}$ the number of
galaxy/void data-random pairs, in a bin at separation $R$.  The total
number of galaxies, voids, galaxy-randoms and voids-randoms are $N_g$,
$N_v$, $N_{rg}$ and $N_{rv}$, respectively. In all cases we generated random catalogues having 10 times the number of galaxies than our data samples.

\medskip

The 1D mock mean void-matter correlation function, $\xi_{v-{\rm
    DM}}(R)$, is displayed in Fig.\ref{Fig1} as the black data points.
We also show the void-halo correlation function (red points), using
voids identified in real-space mocks before applying RSD.  In addition
we compare the same measurements after RSD is applied (dashed lines),
including the 6dFGS measurement.  For clarity we do not show the
errors in the redshift-space measurements, which are similar to the
real-space case.  We see that RSD accentuates the features of the void
profile: it makes the inner density profile steeper and the ridge
higher.
\begin{widetext}
\begin{figure*}[h]
\centering
\begin{tabular}{cc}
\includegraphics[width=\columnwidth]{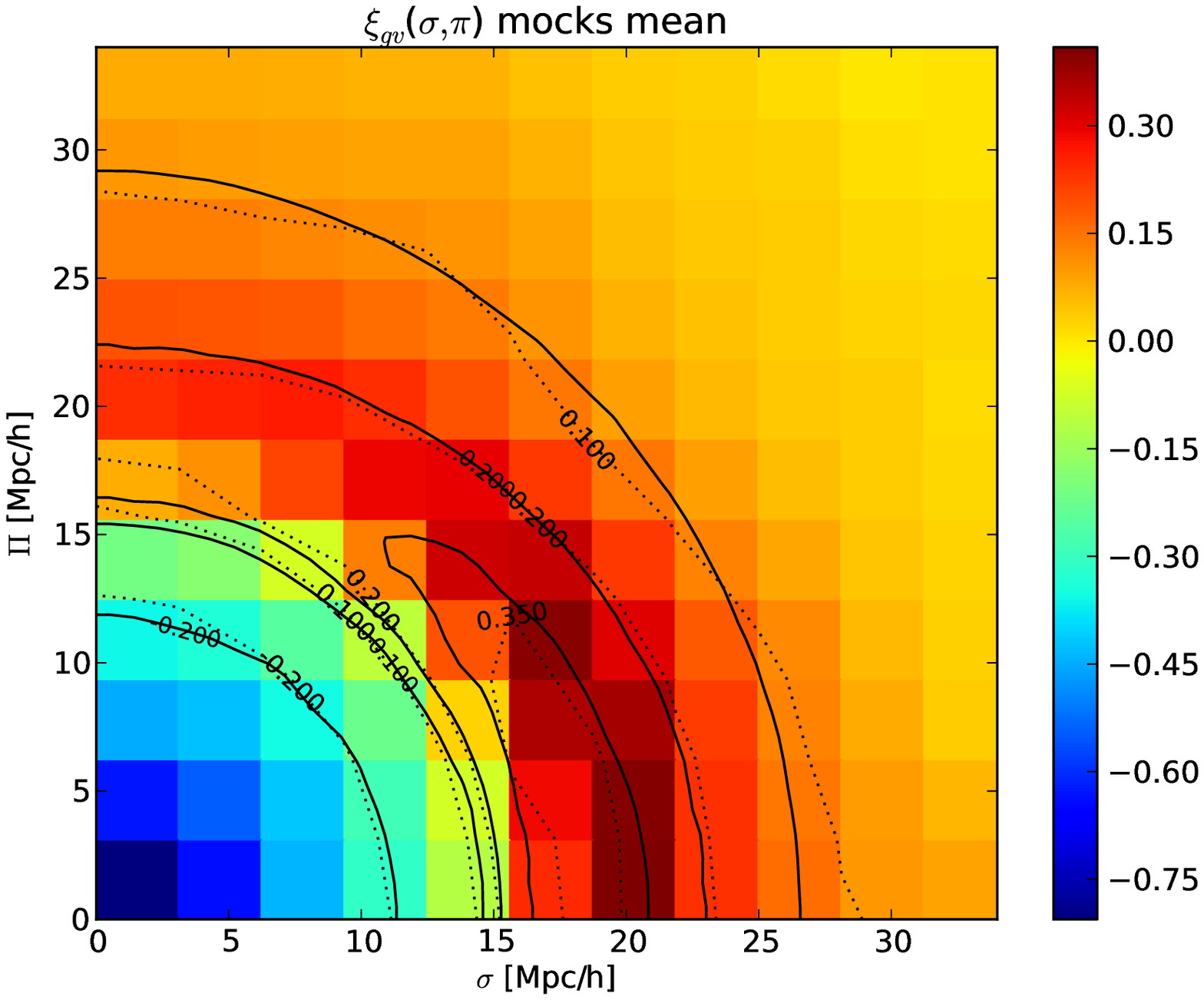} &
\includegraphics[width=\columnwidth]{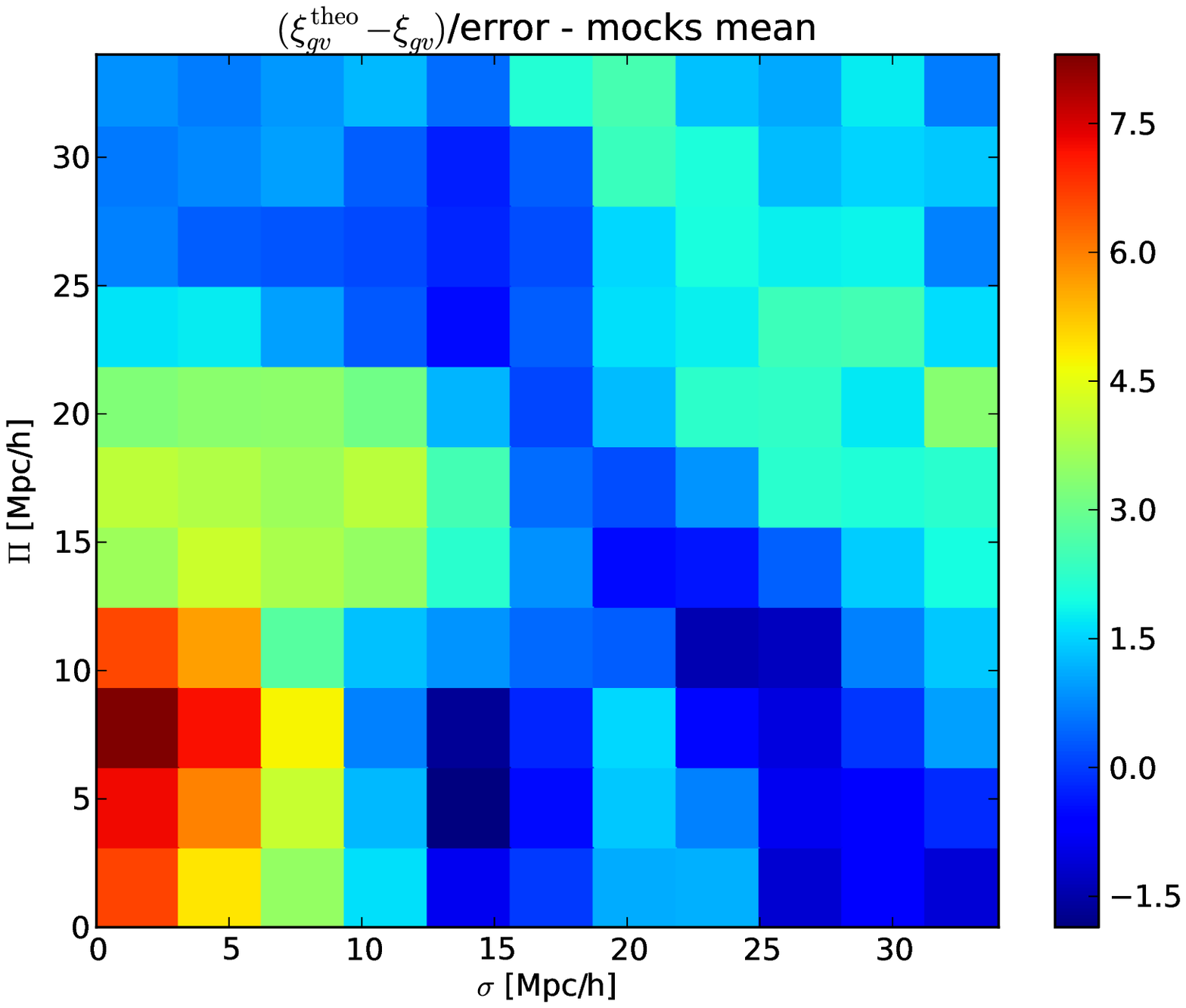} \\
\includegraphics[width=\columnwidth]{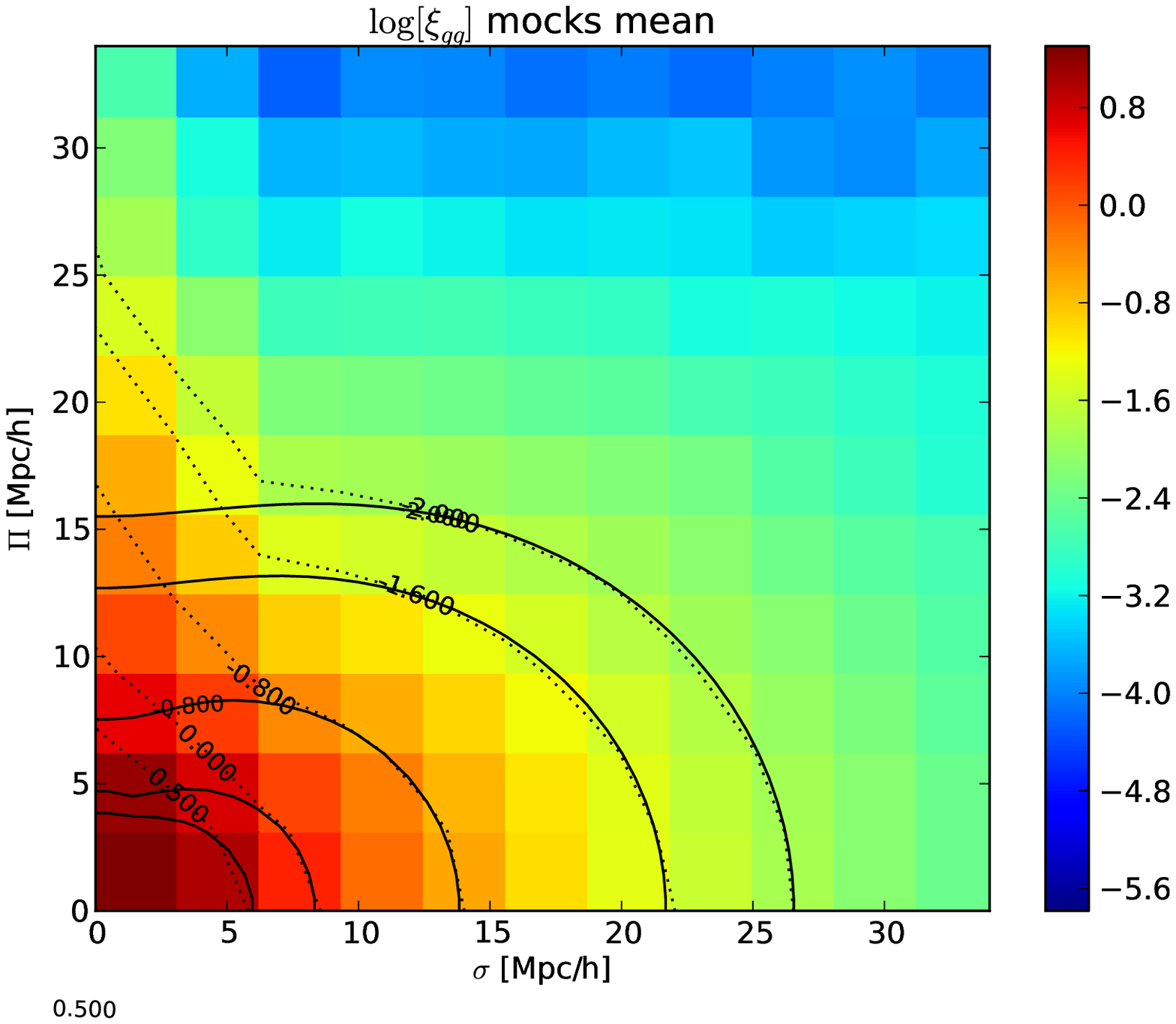} &
\includegraphics[width=\columnwidth]{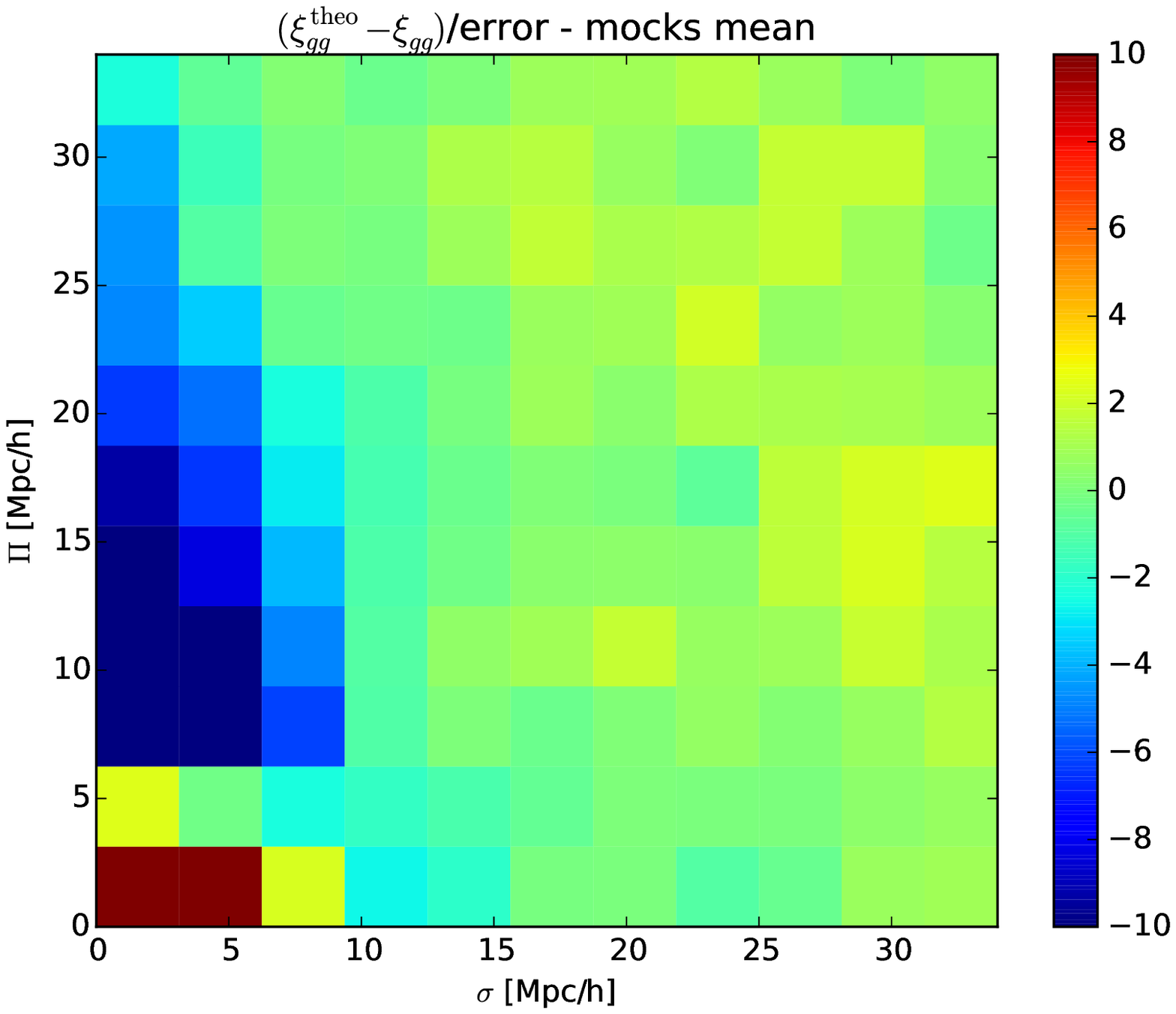}
\end{tabular}
\caption{The mean measurement of the mock 2D void-galaxy correlation
  function (upper-left panel) and galaxy-galaxy correlation function
  (lower-left panel).  The solid lines show the best-fitting model
  assuming a Gaussian pairwise velocity distribution, and the dotted
  lines show iso-contours of the data, noting that the fitting region
  for the galaxy-galaxy correlation function is $\sigma > 7.5 \,
  h^{-1}$ Mpc.  The right-hand panels show the residual between the
  measurements and best-fitting theory in each case, scaled by the
  standard deviation across the mocks. For the galaxy-galaxy residual we impose of min/max cut of -10/+10 in order to distinguish the variations across the right hand side of the plot.  We note that the mock mean is
  a substantially more accurate test of the model than a single
  dataset, and some significant deviations from the model are
  detected.  However, the resulting growth rate fits are unbiased.}
\label{Figmocksxi}
\end{figure*}
\end{widetext}

\subsection*{Model fits to the mock 2D correlation functions}
We computed the 2D void-halo correlation function for Mocks A, and the
halo-halo correlation functions for both Mocks A \& B, using the LS
estimator of Eq.\ref{LSeq}. Indeed, it is interesting to also measure the galaxy-galaxy correlation in mocks A in order to: (i) test if the inferred linear bias is the same as the one inferred from the galaxy-void measurement. (ii) Confirm that the inferred value of the growth rate is the same as the one inferred from the galaxy-void clustering. In fact, there should be a limit of the void size where non-linear effects should impact the value of the growth rate inside large voids. Hence we checked that this effect does not occur for our selected voids by testing the consistency of the growth rate within the same mocks A. When measuring the correlation function for Mocks B, which include the varying survey selection function, we
used minimum variance weights \cite{FeldmanKP,Beutler6dF}
\begin{equation}
w_i = \frac{1}{1 + n_i \, P_0} ,
\end{equation}
where (following \cite{Beutler6dF}) $P_0 = 1600 \, h^{-3}$ Mpc$^3$ and
$n_i$ is the galaxy number density at the location of the $i^{\rm th}$
object.  In Eq.\ref{LSeq}, the ratio of random objects to data objects
then becomes

\begin{equation}
\frac{N_{rg}}{N_g} \rightarrow \frac{\sum_{i=1}^{N_{rg}} w_i}{\sum_{j=1}^{N_g} w_j} .
\end{equation}
We computed the 2D correlation functions of 20 mocks in $(\sigma,\pi)$
bins of width $3 \, h^{-1}$ Mpc in the range $0 - 54 \, h^{-1}$ Mpc,
and used these measurements to construct the standard deviation in
each bin, $\sigma_{\rm mocks}$.

\medskip

For our first analysis we fitted the model to the \textit{mock mean 2D
  correlation function}, with an error in each bin given by $\Delta
\xi = \sigma_{\rm mocks}/\sqrt{N_{\rm mocks}}$.  This allows us to
perform precise systematics tests of Eq.\ref{theogg},\ref{theovg},
using a mock dataset with a statistical error far smaller than the
real 6dFGS dataset.

\medskip

At small scales the galaxy-galaxy correlation function is dominated by
the FoG effect, which can not be described by the linear theory and
pairwise velocity dispersion models of Eq.\ref{theogg}.  Therefore,
small $\sigma$ bins are often excluded when computing
the $\chi^2$ (Eq.\ref{1Dchisq}).  For these reasons we apply a cut $\sigma_{cut} >
7.5 \, h^{-1}$ Mpc when fitting the galaxy-galaxy correlation
function, while we keep all the separation bins for the void-galaxy
correlation function.  We consider below the sensitivity of our
results to these choices.

\medskip

We performed our fit using a Metropolis-Hastings \textit{Markov Chain
  Monte Carlo} (MCMC) analysis for the parameters $\Theta =
(f\sigma_8, b\sigma_8, \sigma_v)$, analyzing our Monte Carlo chains
using the module \textit{GetDist} developed by A. Lewis \cite{Lewis}.
We used priors $f\sigma_8 = [0.02,0.71]$, $b\sigma_8 = [0.4,1.58]$ and
$\sigma_v = [25,600]$ km s$^{-1}$, although our results are not
sensitive to these choices.  We computed the likelihood of each model
assuming
\begin{equation}
\chi^2(\Theta) = \sum_{\sigma,\pi} \left[ \frac{\xi^{\rm
      data}(\sigma,\pi) - \xi^{\rm theo}(\Theta,\sigma,\pi)}{\rm
    \Delta \xi (\sigma,\pi)} \right]^2 .
\label{1Dchisq}
\end{equation}

\medskip

We cannot numerically determine the large covariance matrix between
different $(\sigma,\pi)$ bins sufficiently accurately to allow it to
be inverted when determining the $\chi^2$ statistic, so in
Eq.\ref{1Dchisq} we assumed no correlation between bins.  Our MCMC fit
will therefore not produce robust parameter errors, and we instead
used the \textit{dispersion of the best-fitting parameter values
  between individual mocks}, $\sigma_{IM}$, as a more accurate
estimate of the resulting errors.  This scatter, which is typically
double the parameter error obtained by the MCMC, naturally includes
the effect of data correlations.  When fitting to the mock mean, we
report a scaled parameter error $\sigma_{IM}/\sqrt{N_{\rm mocks}}$.

\medskip

We report the best-fitting parameter values and errors of our fits to
the mock mean galaxy-galaxy and void-galaxy correlation functions, and
minimum $\chi^2$ values, in Tab.\ref{Tab1}.  We find that both the
Gaussian and the Lorentzian models lead to similar constraints on the
growth rate, and that the best-fitting growth rates are consistent
with the fiducial cosmology of the mocks ($f\sigma_8 = 0.26^{0.55}
\times 0.79 \sim 0.38$), validating our models.  The fits to Mocks A
show that the fiducial growth rate is recovered around both voids and
galaxies for a consistent tracer population, and that our choice of
void size produces no unwanted systematic effect due to non-linearity
or inhomogeneity.  The best-fitting $\chi^2$ values are high for both
statistics, although we note that these values neglect the
off-diagonal elements of the covariance matrix, and that the mock mean
provides a far more precise diagnostic of systematics than the real
survey data.

\medskip

The galaxy-galaxy RSD provides weaker constraints on $\sigma_v$ than
the void-galaxy correlation function, due to our exclusion of small
$\sigma$ scales from the fit in this case.  The error in $\sigma_v$ is
sensitive to this cut, as we will see in Fig.\ref{Figchisqcv}.
Tab.\ref{Tab1} also lists best-fitting values for the galaxy bias
factor.  We note that the galaxy bias factor for Mocks B is
significantly higher than for Mocks A, because of the selection of
more massive halos required to match the 6dFGS sample at higher
redshifts, and the upweighting of those halos by the FKP weights.  The
comparison of the results of the void-galaxy and galaxy-galaxy
correlation function fits for Mocks A allows us to verify that the
measured tracer bias is consistent in the two cases, implying that
there is not a environmental dependence of this parameter.

\medskip
\begin{widetext}
\begin{tiny}
\begin{table*}[t]
\centering
\begin{tabular}{l c c c c c c c c c c c}
 & Mocks & $b\sigma_8$ & $\rm \sigma_{IM} $ & $\; f\sigma_8$ & $\rm \sigma_{IM} $& $ \sigma_v [km.s^{-1}]$  & $\rm \sigma_{IM} $& $\chi^2/d.o.f$  \\ 
\hline \hline
$\xi_{gg}$ $L$ & A &$ 0.66$  &  $\pm 0.02$   &  $\; 0.37$ &  $\pm 0.03$  &$134$ &  $\pm 21$ & $497/192$ \\
$\xi_{vg}$ $L$ & A &$ 0.67$  &  $\pm 0.01$   &  $\; 0.38$ &  $\pm 0.02$  &$126$ & $\pm 8.5$ & $920/277$ \\
$\xi_{gg}$ $G$ & A &$ 0.66$  &  $\pm 0.02$  &  $\; 0.37$ & $\pm 0.03$  &$118$ &  $\pm 19$ & $497/192$ \\
$\xi_{vg}$ $G$ & A &$ 0.67$  &  $\pm 0.01$   &  $\; 0.38$ &  $\pm 0.02$  &$122$ & $\pm 9$ & $925/277$ \\
$\xi_{gg}$ $G$ & B &$ $1.01  &  $\pm 0.01$  &  $\; 0.38$ &  $\pm 0.01$  &$102$ &  $\pm 21$ & $327/192$ \\
$\xi_{gg}$ $L$ & B &$ $1.00  & $\pm 0.01$  &  $\; 0.38$ &  $\pm 0.01$  &$100$ &  $\pm 25$ & $326/192$ \\
\hline
\end{tabular}
\caption{Parameter constraints obtained from fitting to the mock mean
  2D galaxy-galaxy correlation function $\xi_{gg}$ and void-galaxy
  correlation function $\xi_{vg}$ for Mocks A and B, assuming
  Lorentzian ($L$) and Gaussian ($G$) models for the pairwise velocity
  dispersion.  The reported parameter errors are the scatter in the
  fits to individual mocks, scaled by $\sqrt{N_{\rm mocks}}$.  The
  $\chi^2$ values are derived from the MCMC fit to the mock mean,
  which is impacted by neglecting off-diagonal covariance. The fiducial cosmology in the mocks is $f\sigma_8 = 0.26^{0.55}
\times 0.79 \sim 0.38$.}
\label{Tab1}
\end{table*}
\end{tiny}
\end{widetext}
\begin{widetext}
\begin{tiny}
\begin{table*}[t]
\centering
\begin{tabular}{l c c c c}
 &Mocks& $\bar{\sigma_8 b}/\sigma_{\sigma_8 b} $ & $\bar{f\sigma_8}/\sigma_{f\sigma_8}$ & $ \bar{\sigma_v}/\sigma_{\sigma_v} [km.s^{-1}]$ \\ 
\hline \hline
$\xi_{gg}$ $L$ & A&$0.58/0.20$ & $0.36/0.24$ & $293/166$ \\
$\xi_{vg}$ $L$ & A&$0.70/0.08$ & $0.44/0.18$ & $164/68$ \\
$\xi_{gg}$ $G$ & A&$0.58/0.19$ & $0.38/0.29$ & $270/150$ \\
$\xi_{vg}$ $G$ & A&$0.70/0.08$  & $0.42/0.19$ & $181.5/72$ \\
$\xi_{gg}$ $G$ & B&$1.0/0.06$ & $0.39/0.06$ & $111/92$ \\
$\xi_{gg}$ $L$ & B&$1.0/0.06$ & $0.39/0.06$ & $125/113$ \\
\hline
\end{tabular}
\caption{Parameter constraints obtained by fitting to each individual
  mock and measuring the resulting mean and standard deviation of the
  best-fitting parameters, for the 2D galaxy-galaxy correlation
  function $\xi_{gg}$ and void-galaxy correlation function $\xi_{vg}$,
  assuming Lorentzian ($L$) and Gaussian ($G$) models for the pairwise
  velocity dispersion. The fiducial cosmology in the mocks is $f\sigma_8 = 0.26^{0.55}
\times 0.79 \sim 0.38$}
\label{Tab1bis}
\end{table*}
\end{tiny}
\end{widetext}

In Tab.\ref{Tab1bis} we report summary statistics of the fits of our
model to the individual mock catalogues, listing the mean values of
the best-fitting parameters ($\bar{f \sigma_8}$, $\bar{b \sigma_8}$,
$\bar{\sigma_v}$) and their dispersion across the mock catalogues
($\sigma_{f\sigma_8}$, $\sigma_{b\sigma_8}$, $\sigma_{\sigma_v}$).
The mean values are consistent with the best fit to the mock mean,
indicating that our approach is unbiased.
\medskip

We checked the dependence of the best-fitting parameter values on the
range of scales included in our analysis.  In the upper panel of
Fig.\ref{Figchisqcv}, we show the variation of the best-fitting values
with the cutting scale $\sigma_{cut}$, for the fits to the
galaxy-galaxy correlation function of Mocks B.  The triangles (red for
model $G$ and orange for model $L$) show the result from fitting to
the mock mean, while the unfilled circles correspond to the mean
parameter fit to the individual mocks.  The minimum reduced $\chi^2$
is shown in the bottom panel.  Deviations are seen when including the
first bin, which we expect to be most strongly affected, although our
results do not show a strong dependence on $\sigma_{cut}$ and we adopt
a baseline $\sigma_{cut} = 7.5 \, h^{-1}$ Mpc for our analyses.
\medskip

A similar analysis of the void-galaxy correlation function of Mocks A
is shown in the lower panel of Fig.\ref{Figchisqcv} where, given the
absence of non-linear pairwise velocities, we now consider a cut as a
function of the total separation, $R_{cut} = \sqrt{\pi^2+\sigma^2}$.
This is motivated by the possibility that linear theory may break down
at the centre of the voids where $\delta_v(R \rightarrow 0) \approx
-1$ \cite{Caietal2016}.  We plot the best-fitting parameters as a
function of $R_{cut}$ as well as the reduced $\chi^2$ for model $G$
(blue lines) and model $L$ (cyan lines).  The fits to the mock mean
are shown by the triangles, while the unfilled circles correspond to
the mean parameter fit to the individual mocks.  In this case, we find
a low sensitivity of the results to the value of $R_{cut}$.  The
best-fitting parameters are consistent with our fiducial cosmology
when we use all scales $(R_{cut} = 0)$ in Eq.\ref{1Dchisq}.

\section{Application to 6dFGS}
\label{sec4}
\begin{widetext}
\begin{figure*}[t]
\centering
\begin{tabular}{cc}
\includegraphics[width=\columnwidth]{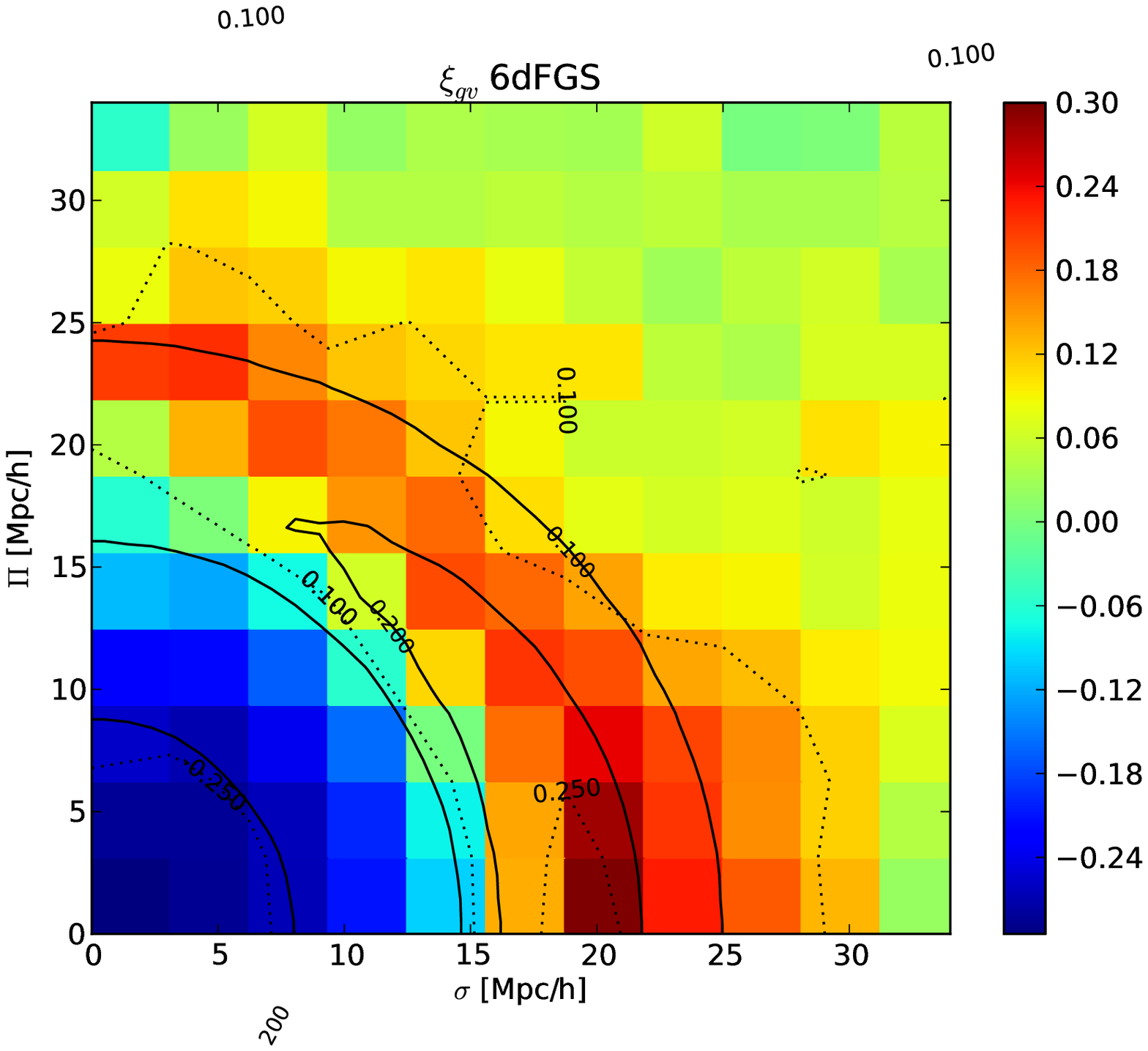} &
\includegraphics[width=\columnwidth]{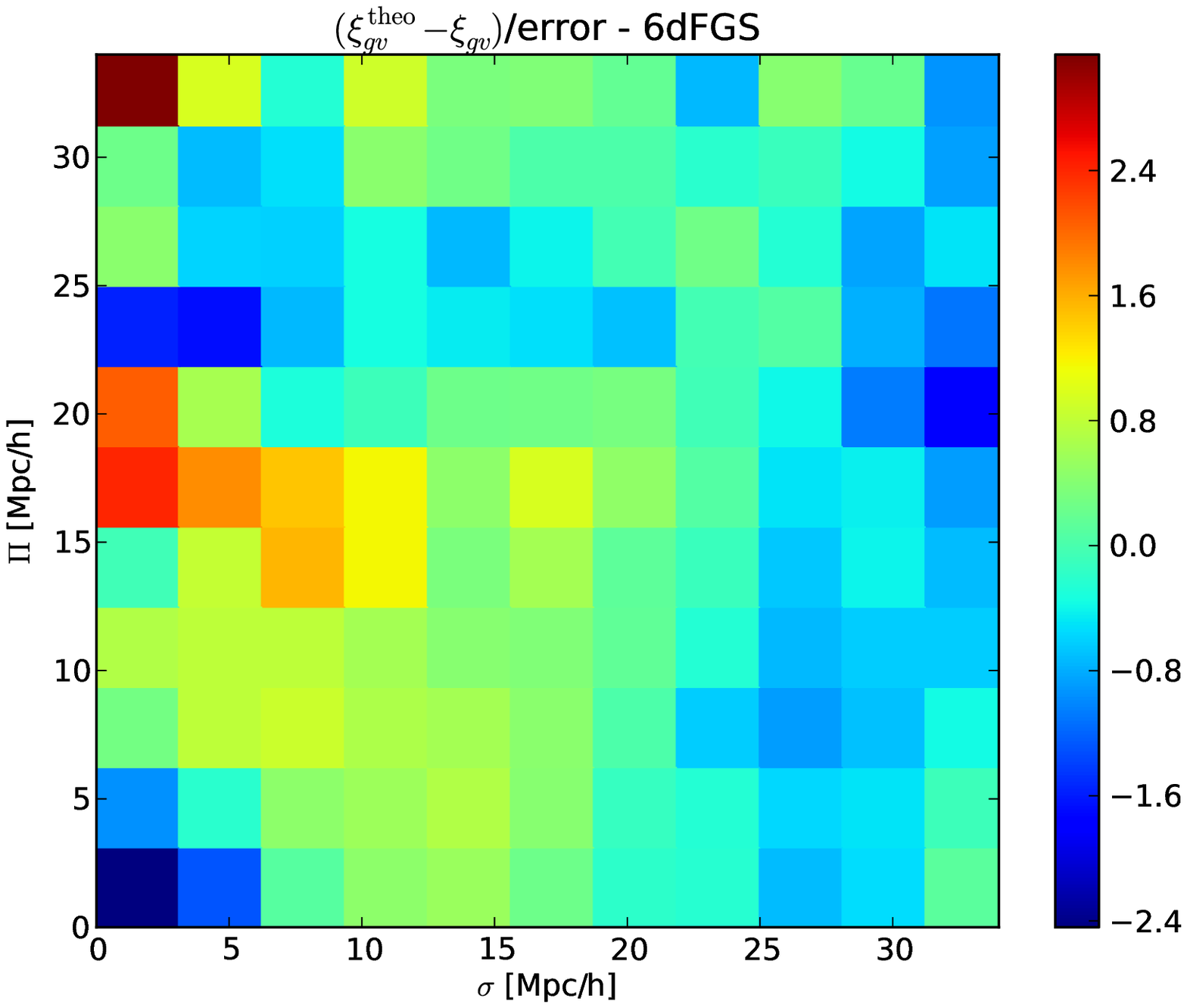}\\
\includegraphics[width=\columnwidth]{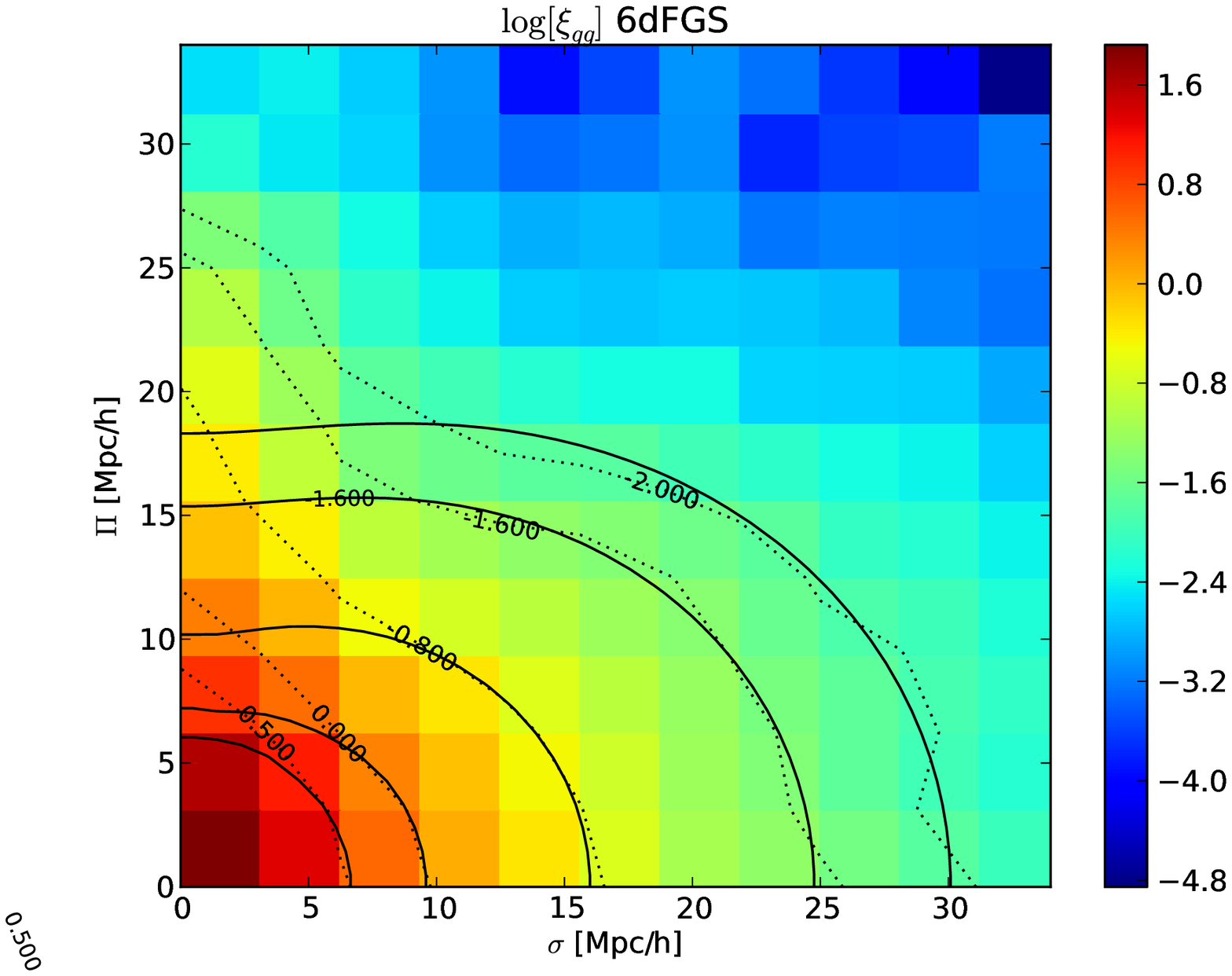} &
\includegraphics[width=\columnwidth]{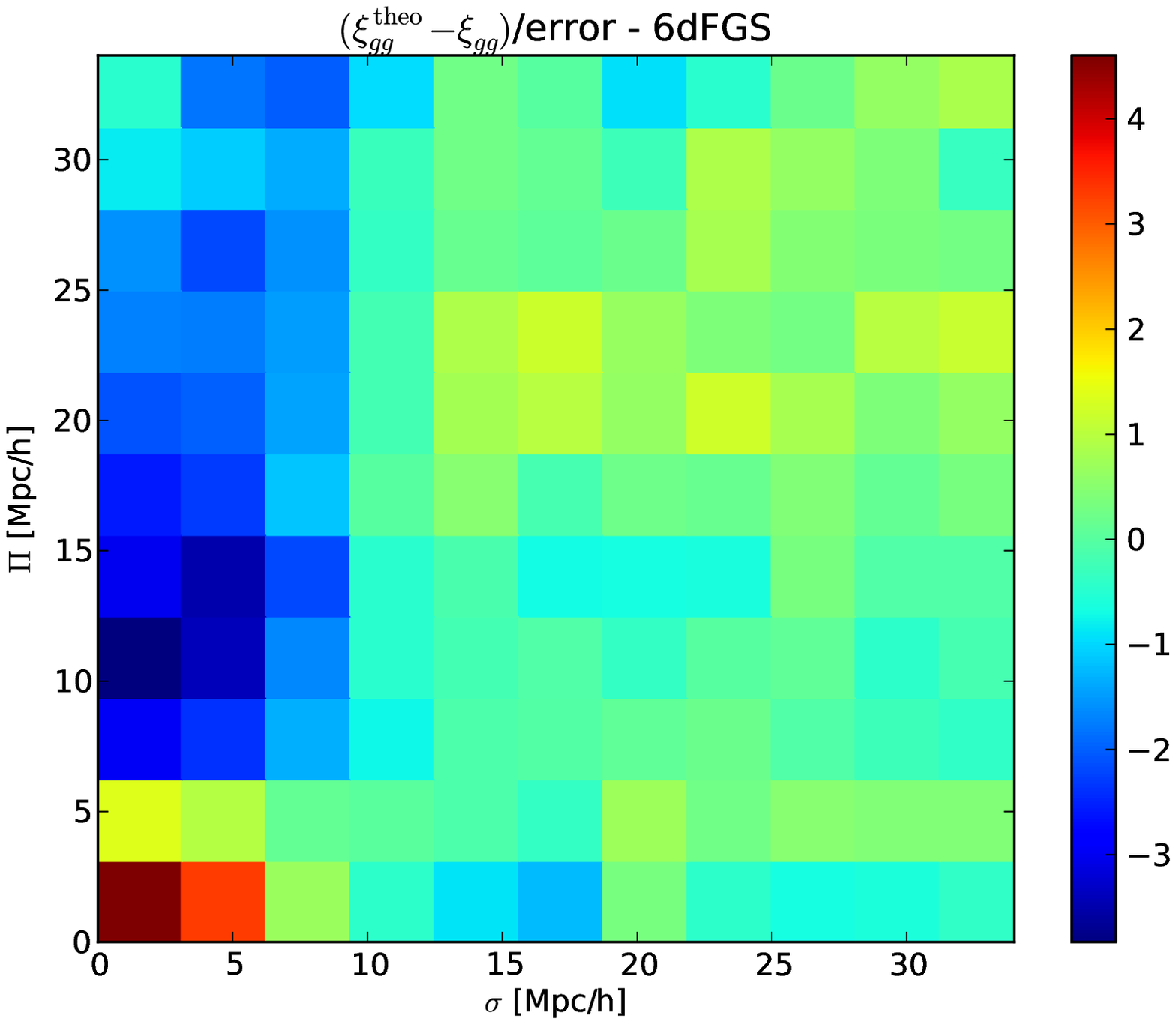}
\end{tabular}
\caption{The 2D void-galaxy correlation function (upper left panel)
  and galaxy-galaxy correlation function (lower left panel) of the
  6dFGS dataset.  The solid lines show the best-fitting model assuming
  a Gaussian pairwise velocity dispersion, and the dotted lines show
  iso-contours of the data, noting that the fitting region for the
  galaxy-galaxy correlation function is $\sigma > 7.5 \, h^{-1}$ Mpc.
  The right-hand panels show the corresponding residual between the
  measurement and best-fitting model, scaled by the error in each bin.
  In general, there are not significant residuals within the fitted
  region.}
\label{Figxi2d}
\end{figure*}
\end{widetext}
\begin{widetext}
\begin{figure*}[h]
\subfloat[Galaxy-galaxy fits as a function of the cutting scale]{\includegraphics[scale=0.4]{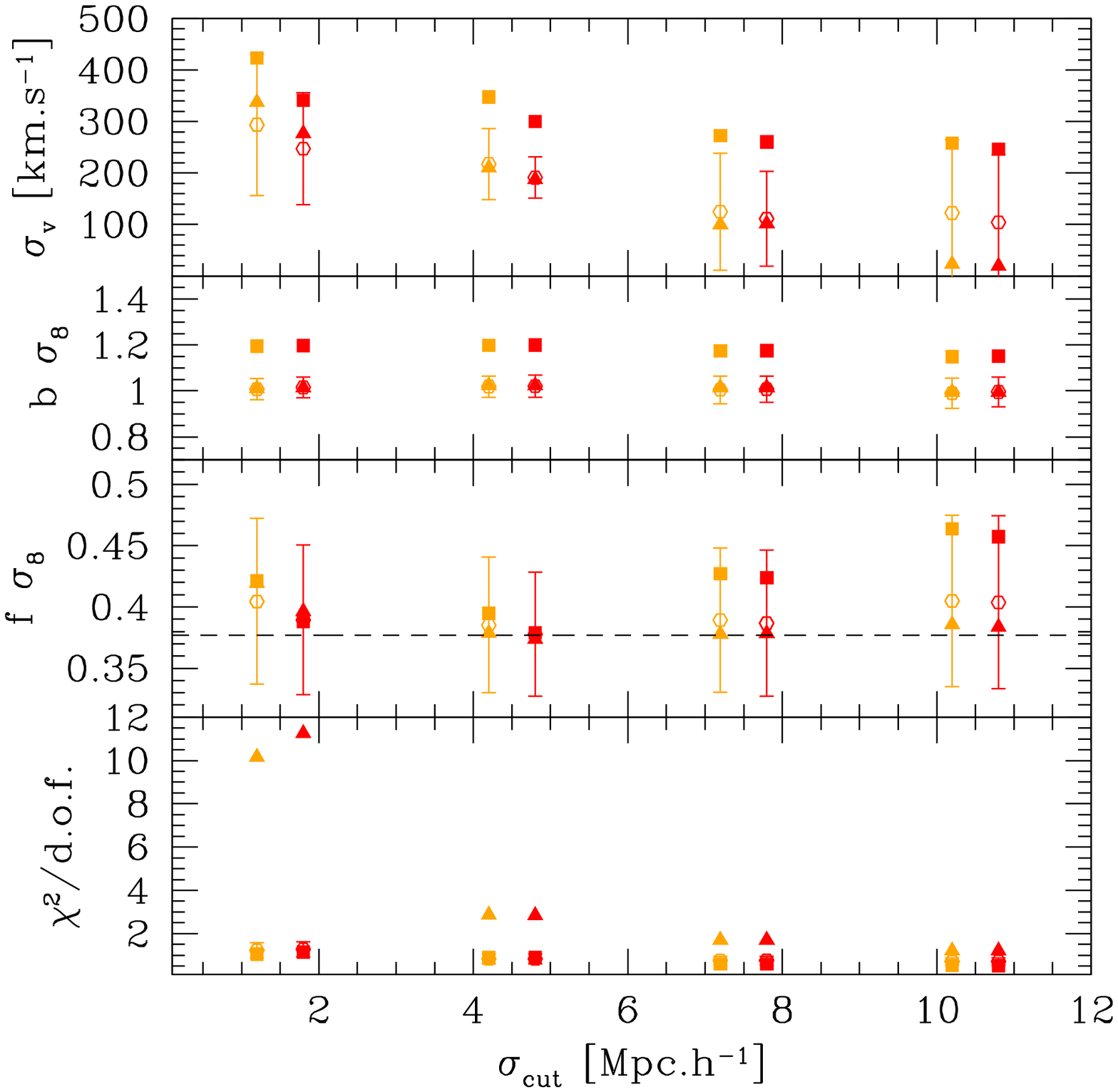}} 
\subfloat[Void-galaxy fits as a function of the cutting scale]{\includegraphics[scale=0.4]{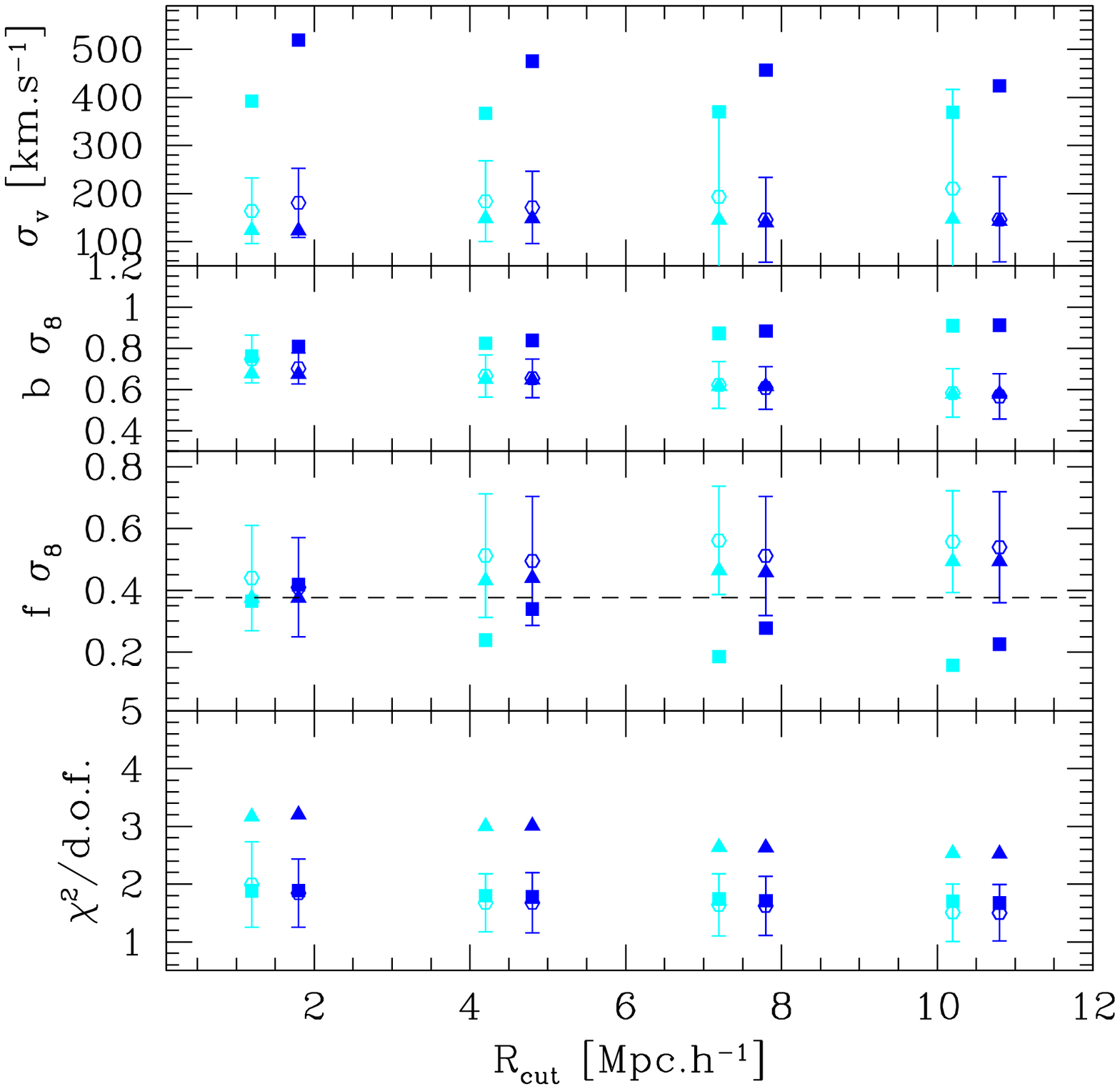}}
\caption{The influence of the fitting range on parameter fits to the
  galaxy-galaxy (left panel) and void-galaxy (right panel) correlation
  functions.  In both cases we show the result using the Gaussian
  pairwise velocity model (blue and red) and the Lorentzian model
  (orange and light blue).  The squares correspond to the 6dFGS
  constraints, the triangles correspond to the fits to the mock mean,
  and the non-filled circles correspond to the mean of the fits to
  individual mocks, with the error bars as the standard deviation.  We
  offset points along the $x$-axis for clarity.}
\label{Figchisqcv}
\end{figure*}
\end{widetext}

\subsection*{Galaxy and void samples}

The 6dF Galaxy Survey was undertaken with the multi-fibre instrument
on the UK Schmidt Telescope between 2001 and 2006. The median redshift
of the survey is $z=0.052$ and it covers nearly the entire southern
sky.  A full description of the survey can be find in
\cite{Jones2004,Jones2006} including comparisons between 6dFGS, 2dFGRS
and SDSS.  In this analysis we utilized the same $K$-band selected
6dFGS sub-sample, consisting of $\sim 70500$ galaxies, as constructed
for the analysis of the baryon acoustic peak by \cite{BeutlerBAO}.  We
also used random catalogues following the same angular and redshift
selection as the data sample, generated by \cite{BeutlerBAO}.
\medskip

We constructed different 6dFGS sub-samples for analyzing the
galaxy-galaxy and void-galaxy correlation functions.  For the
measurement of the void-galaxy correlation function, we first
constructed a volume-limited catalogue corresponding to an
approximately constant number density.  This step is crucial in order
to apply our measurement of the 1D real-space void-matter correlation
function in Eq.\ref{xibias}, and to avoid any evolution in the void
properties with redshift.  We built the volume-limited catalogue by
determining the absolute magnitude $M$ of each galaxy using
\begin{equation}
m - M = 5 \log_{10}{D_L(z)} + 25 + K(z) ,
\label{Magnieq}
\end{equation}
where $m$ is the apparent $K$-band magnitude, $D_L(z)$ is the
luminosity distance in Mpc and $K(z)$ is the $K$-correction
\cite{0210394v1,Mannucci2001}.  For this analysis we set the maximum
redshift of the sample to $z_{\rm max} = 0.05$, in order to obtain a
sample with a sufficiently high number density.  The faint magnitude
limit of the survey is $m_{\rm faint} = 12.75$, and we selected all
galaxies brighter than $M_{\rm faint}$ in the redshift range $z <
z_{\rm max}$, where $M_{\rm faint}$ is computed from Eq.\ref{Magnieq}
with $z = z_{\rm max}$.  We identified voids in the catalogue using
the algorithm described in Sec.\ref{sec3}, leading to the
identification of $\sim 1400$ voids.

\subsection*{Measurement of the correlation function}
We transformed the angular co-ordinates and redshifts of the galaxies
to co-moving Cartesian coordinates assuming the same fiducial
cosmology as our mock catalogue $(\Omega_m=0.26)$, although we note
that the Alcock-Paczynski effect is negligible at low redshift.  The
separation of two galaxies along the line of sight $\pi$ and across
the line of sight $\sigma$ is measured in the same manner as Mocks B
using
\begin{equation}
\begin{split}
&\pi=\frac{\|\mathbf{s}.\mathbf{h}\|}{\|\mathbf{s}\|}\\
&\sigma=\sqrt{\|\mathbf{h}\|^2-\pi^2} ,
\end{split}
\end{equation}
where $\mathbf{h} = \mathbf{s_1} - \mathbf{s_2}$ is the separation of
the galaxies in redshift space and $\mathbf{s} = (\mathbf{s_1} +
\mathbf{s_2})/2$ is the mean distance to the galaxy pair.
\medskip

Fig.\ref{Figxi2d} displays the measured 2D galaxy-galaxy correlation
function (lower left) and void-galaxy correlation function (lower
right) for the 6dFGS dataset.  For the galaxy-galaxy correlation
function, we can see the elongation at small scales along the line of
sight (FoG), due to the random motion of galaxies within halos.  On
larger scales, we observe the Kaiser effect due to coherent bulk
flows.  For the void-galaxy correlation function, we can detect an
apparent asymmetry within the void ($< 15 \, h^{-1}$ Mpc): the
`emptiness' is larger along the line of sight due to the cosmic
expansion, and the ridge of the void ($\sim 20 \, h^{-1}$ Mpc) tends
to be erased due to the velocity dispersion.  The Kaiser effect can
also be observed: the signal is enhanced across the line of sight,
especially on the ridge.
\medskip

We obtained the error in the 6dFGS void-galaxy and galaxy-galaxy
correlation functions using the dispersion in the measurements from
Mocks A and B, respectively.  We scaled the standard deviation of the
void-galaxy mock measurements to allow for the slightly different
volumes of Mock A and the real dataset:
\begin{equation}
\Delta \xi = \sqrt{\frac{V_{\rm mock}}{V_{\rm 6dFGS-cut}}} \times \sigma_{mock}
\end{equation}
where $V_{\rm 6dFGS-cut} \sim 179^{3} \, h^{-3}$ Mpc$^3$ and the
scaling factor is $0.64$.  The parameter errors are also scaled by
this correction factor.  No volume-scaling is needed for the
galaxy-galaxy correlation functions, since Mocks B sample the exact
survey selection function.

\subsection*{Growth rate measurement in different environments}
\label{secresult}

We fitted our RSD model to the 6dFGS data using the MCMC pipeline
described in Section \ref{sec3}.  As previously discussed, we obtain
robust parameter errors using the dispersion of the fits to the mock
catalogues.
\medskip

We report the best-fitting parameter values and their errors in
Tab.\ref{Tab2}.  Our measurement of the growth rate for the average of
models L and G is $f \sigma_8 = 0.42 \pm 0.06$ for the galaxy-galaxy
RSD and $f \sigma_8 = 0.39 \pm 0.11$ for the void-galaxy RSD.  We
observe larger uncertainties in the growth rate measured using the
void-galaxy correlation function, although the two measurements are
consistent within the statistical errors.  The minimum $\chi^2$
values, also listed in Tab.\ref{Tab2}, are lower than those found for
the more accurate mock mean dataset, but we note that they are still
impacted by the assumption of a diagonal covariance matrix.  The
right-hand panels of Figure \ref{Figxi2d} show the residuals between
the data and best-fitting models.  Our measurement is in very good
agreement with the previous 6dFGS galaxy-galaxy RSD analysis
\cite{Beutler6dF}, which obtained $f \sigma_8 = 0.42 \pm 0.05$.
\medskip

The difference in the best-fitting bias parameters for $\xi_{gg}$ and
$\xi_{vg}$ is due to the different galaxy samples used: for the
galaxy-galaxy analysis we adopt a flux-limited sample across a wider
redshift range, and up-weight more luminous, highly-biased galaxies.
The best-fitting bias values are comparable with those found in the
corresponding mock catalogue analyses in each case, although some
differences remain.
\medskip

These results are obtained with a cut $\sigma_{cut} = 7.5 \, h^{-1}$
Mpc for the galaxy-galaxy correlation function, and using all bins for
the void-galaxy correlation function.  This is motivated by the
mock-catalogue analysis and the lack of sensitivity of our
best-fitting parameters to these choices, which is illustrated by
Fig.\ref{Figchisqcv}.  For $\sigma_{cut} > 4.5 \, h^{-1}$ Mpc, the
goodness-of-fit and best-fitting parameters do not significantly
change for the galaxy-galaxy correlation function (left panel),
independently of the model (see the red/orange solid lines).  The
best-fitting $\chi^2$ of the void-galaxy correlation function (right
panel) remains unchanged at all scales, independently of the model
(see the blue/light blue solid lines).
\medskip

Overall, the growth rate measurements are consistent between the
void-galaxy and galaxy-galaxy RSD.  One might think about combining
these measurements to improve the uncertainties.  However we do not
expect a significant improvement since the growth uncertainties from
the void-galaxy RSD are double those of the galaxy-galaxy RSD, and the
measurements are correlated.  Hence, the novelty of our result relies
on the comparison of the growth between different environments.

\begin{widetext}
\begin{tiny}
\begin{table*}[t]
\centering
\begin{tabular}{l c c c c c c c c c c}
 & $b\sigma_8$ & $\rm \sigma_{IM} $ & $\; f\sigma_8$ & $\rm \sigma_{IM} $& $ \sigma_v [km.s^{-1}]$ & $\rm \sigma_{IM} $& $\chi^2/d.o.f$  \\ 
\hline \hline
$\xi_{gg}$ $L$ & $ 1.17$   & $\pm 0.06$   &  $\; 0.43$ & $\pm 0.06$  &$273$ & $\pm 92$ & $114/192$ \\
$\xi_{vg}$ $L$ & $0.76$   & $\pm 0.05$   &  $\; 0.36$ & $\pm 0.11$  &$390$  & $\pm 43$ & $530/289$ \\
$\xi_{gg}$ $G$ & $1.17$  & $\pm 0.06$  &  $\; 0.42$  & $\pm 0.06$  &$261$  & $\pm 113$ & $116/192$ \\
$\xi_{vg}$ $G$ & $ 0.80$ & $\pm 0.05$   &  $\; 0.43$ & $\pm 0.12$  &$515$ & $\pm 46$ & $536/289$ \\
\hline
\end{tabular}
\caption{Parameter constraints obtained from fitting to the 6dFGS 2D
  galaxy-galaxy correlation function $\xi_{gg}$ and void-galaxy
  correlation function $\xi_{vg}$, assuming Lorentzian ($L$) and
  Gaussian ($G$) models for the pairwise velocity dispersion.  We
  determine the parameter errors using the standard deviation of the
  parameter fits to individual mocks.}
\label{Tab2}
\end{table*}
\end{tiny}
\end{widetext}

\section{Conclusion}
\label{sec5}

In this work we provide the first direct comparison of the cosmic
growth rate measured in two different environments of the same galaxy
survey, by fitting to Redshift Space Distortion in the galaxy-galaxy
and void-galaxy correlation functions of the 6-degree Field Galaxy
Survey.  As a low-redshift survey, our 6dFGS measurements are
particularly relevant for probing the late-time domination of dark
energy, and are insensitive to the Alcock-Pacynski effect.  We find
voids using a new void-finder which identifies under-densities
matching supplied density profile criteria \cite{IA_voidfinder}.  We
also note that our measurement of the growth using RSD around voids is
the first performed at low redshift and in the southern hemisphere.

\medskip

We determine similar growth rate measurements around galaxies
($f\sigma_8 = 0.42 \pm 0.06$) and $\sim 20 \, h^{-1}$ Mpc
underdensities ($f\sigma_8 = 0.39 \pm 0.11$), finding no evidence of
an environmental dependence of gravitational physics.  We validate our
models, and estimate the errors in our measurements, using mock galaxy
catalogues.  Extracting the complementary cosmological information
present in different environments \citep{AchitouvBlake2015,
  Kitaura2015} will be a powerful test of physics for both current
galaxy redshift surveys and future projects such as \textit{Euclid}
\cite{EUCLIDsurvey}.

\medskip

Our analysis could be extended in several ways: direct measurements of
peculiar velocities using standard-candle indicators could further
constrain their radial profile around voids; combining our results
with analyses of other data sets such as SDSS \cite{SDSSsurvey} and
GAMA \cite{Gamasurvey} can probe these effects as a function of
redshift; and a comparison of our measurements with the predictions of
non-standard cosmological models, in particular modified gravity and
interacting dark energy models, would place new constraints on those
frameworks.
 
\section*{Acknowledgments}

We are very grateful to Yann Rasera for facilitating the access to the DEUS N-body simulations. 

The 6dF Galaxy Survey was made possible by contributions from many
individuals towards the instrument, the survey and its science. We
particularly thank Matthew Colless, Heath Jones, Will Saunders, Fred
Watson, Quentin Parker, Mike Read, Lachlan Campbell, Chris Springob,
Christina Magoulas, John Lucey, Jeremy Mould, and Tom Jarrett, as well
as the dedicated staff of the Australian Astronomical Observatory and
other members of the 6dFGS team over the years.

Part of this research was conducted by the Australian Research Council
Centre of Excellence for All-sky Astrophysics (CAASTRO), through
project number CE110001020. We also acknowledge support from the DIM
ACAV of the Region Ile-de-France.

\end{document}